\documentclass[manuscript=article]{achemso}
\setkeys{acs}{usetitle=true}

\usepackage{amsmath}
\usepackage[version=3]{mhchem}
\usepackage{makecell}
\usepackage{multirow}
\usepackage{natbib}
\usepackage{graphicx}
\usepackage{float}
\usepackage{hyperref}
\title[Short title]{Accelerating Complex Materials Discovery with Universal Machine-Learning Potential-Driven Structure Prediction}

\author{Yuqi An}
\affiliation[CityU]{Department of Materials Science and Engineering, City University of Hong Kong, Hong Kong SAR, 999077, China}
\author{Zhenbin Wang}
\affiliation[CityU]{Department of Materials Science and Engineering, City University of Hong Kong, Hong Kong SAR, 999077, China}
\alsoaffiliation[CityU]{School of Energy and Environment, City University of Hong Kong, Hong Kong SAR, 999077, China}
\email{zwan22@cityu.edu.hk}
\date{}

\begin{document}

\maketitle

\begin{abstract}
Universal machine-learning interatomic potentials (uMLIPs) have become powerful tools for accelerating computational materials discovery by replacing expensive first-principles calculations in crystal structure prediction (CSP). However, their effectiveness in identifying novel, complex materials remains uncertain. Here, we systematically assess the capability of a uMLIP (i.e., M3GNet) to accelerate CSP in quaternary oxides. Through extensive exploration of the Sr–Li–Al–O and Ba–Y–Al–O systems, we show that uMLIP can rediscover experimentally known materials absent from its training set and identify seven new thermodynamically and dynamically stable compounds. These include a novel polymorph of \ce{Sr2LiAlO4} ($P3_221$) and a new disordered phase, \ce{Sr2Li4Al2O7} ($P\bar{1}$). Furthermore, our results show stability predictions based on the semilocal PBE functional require cross-validation with higher-level methods, such as SCAN and RPA, to ensure reliability. While uMLIPs substantially reduce the computational cost of CSP, the primary bottleneck has shifted to the efficiency of search algorithms in navigating complex structural spaces. This work highlights both the promise and current limitations of uMLIP-driven CSP in the discovery of new materials.

\end{abstract}

\section{Introduction}
The discovery of new materials is a fundamental driver of technological innovation. While this process has traditionally been guided by experimental intuition and serendipity, such methods are often time-consuming and inefficient. In recent decades, computational materials discovery, particularly through crystal structure prediction (CSP), has emerged as a powerful tool.\cite{oganov2019structure,woodley2008crystal} CSP accelerates the discovery process by predicting the most stable atomic arrangements for a given set of elements before synthesis is attempted. The traditional approach to CSP combines global optimization techniques, such as evolutionary algorithms\cite{oganov2006crystal}, with first-principles calculations. However, this method is severely hampered by its immense computational expense and is consequently restricted to small and simple chemical systems, which limits the exploration of the vast chemical space of complex materials where many technologically relevant properties are found.

To overcome the computational bottleneck of first-principles methods, machine-learning interatomic potentials (MLIPs) have been introduced as fast and accurate surrogates.\cite{behler2007generalized,bartok2010gaussian,shapeev2016moment} Early MLIPs required laborious, system-specific training and suffered from poor transferability, limiting their predictive power. While the subsequent integration of active learning strategies improved efficiency, these approaches still face significant computational hurdles for complex systems. More recently, universal MLIPs (uMLIPs) like M3GNet\cite{chen2022universal}, MACE\cite{batatia2022mace}, and ORB-v3\cite{rhodes2025orb} have been developed, trained on vast datasets to achieve high accuracy in predicting energetics across diverse chemical spaces. They have been successfully applied to a range of applications, including the discovery of new battery materials\cite{guo2024Machinelearning,chen2024Accelerating,jeong2024Machinelearningdriven}, the design of complex alloys\cite{zhu2025Accelerating}, the evaluation of catalytic performance\cite{schaaf2023Accurate,edwards2025Exploring}, and the prediction of mechanical properties\cite{rana2025Prediction,liu2025Study}. Although these uMLIPs hold great promise for aiding materials discovery, their practical application and effectiveness for \textit{de novo} crystal structure prediction of new, complex materials have not yet been systematically demonstrated.

In this work, we systematically investigate the capability of uMLIPs in accelerating complex materials discovery. We selected the Sr-Li-Al-O and Ba-Y-Al-O chemical systems as case studies. These complex quaternary spaces are promising for the development of new phosphor materials for solid-state lighting\cite{wang2018mining}, yet their complexity makes them computationally prohibitive for traditional DFT-based crystal structure prediction methods. Furthermore, as few compounds are experimentally known within these systems, they represent high-potential areas for materials discovery. Using these systems, we demonstrate that M3GNet is capable of not only rediscovering experimentally known materials that were not part of its training dataset, but also efficiently identifying new materials with good thermodynamic and dynamic stability. However, our results also underscore the necessity of higher-level first-principles calculations for accurate final-phase stability evaluations. Moreover, the limiting factor in CSP has shifted toward the efficiency of search algorithms in navigating large, complex structural spaces.
\section{Methods}
\subsection{Universal Machine-Learning Interatomic Potential}
To perform a global potential energy surface search for new crystal structures, we employed the universal machine learning interatomic potential (uMLIP), M3GNet-DIRECT\cite{qi2024Robust}, for all energy and force calculations. M3GNet-DIRECT is an improved model retrained from the original M3GNet\cite{chen2022universal} using the DImensionality-Reduced Encoded Clusters with sTratified (DIRECT) sampling strategy on the Materials Project database (version 2021.2.8). For structure relaxation during the crystal structure search, M3GNet was interfaced with the LAMMPS package via a customized implementation\cite{thompson2022lammps,advsc2023}. 
M3GNet was chosen over other, more accurate uMLIPs(e.g., ORB-v3\cite{rhodes2025orb}) as the test case for uMLIP-driven CSP for two main reasons: (i) it represents the first generation of uMLIPs, and (ii) its training dataset is well-documented and traceable. The latter makes it possible to identify experimentally known materials absent from the training set, enabling a rigorous evaluation of M3GNet's extrapolation capabilities. Specifically, the M3GNet-DIRECT variant was utilized, as it consistently identified structures with lower formation energies than the original model—for instance, a 6 meV/atom improvement for the 12-atom \ce{SrLiAlO3} system (see Table S3).

\subsection{Crystal Structure Prediction}
We performed crystal structure prediction (CSP) using the evolutionary algorithm implemented in the USPEX code.\cite{glass2006uspex} Each search began with an initial population of 100 randomly generated structures. Subsequent generations were produced using four variation operators: heredity (50\%), random generation (30\%), permutation (10\%), and atomic mutation (10\%). This combination ensures a robust balance between global exploration and local optimization of the potential energy surface. The M3GNet-calculated enthalpy served as the fitness criterion for structural selection in each generation. A search was considered converged and terminated if the lowest-enthalpy structure remained unchanged for 35 consecutive generations or after a maximum of 80 generations. To ensure robust and reproducible results, each CSP search was performed independently three times.

We conducted crystal structure searches in the Sr–Li–Al–O and Ba–Y–Al–O chemical systems. These two systems were selected for structural exploration because they contain relatively few known quaternary compounds identified experimentally \cite{wang2018mining}, indicating a promising likelihood of discovering new materials. To balance thorough exploration of the polymorphic landscape with computational efficiency, we limited our search to structures containing fewer than 30 atoms per unit cell. Enforcing charge neutrality within this size constraint yielded 17 distinct compositions for the Sr–Li–Al–O system and 29 for the Ba–Y–Al–O system. These compositions were determined using the standard valence states of each element: Sr (+2), Ba (+2), Li (+1), Al (+3), Y (+3), and O (–2).

\subsection{Density Functional Theory Calculations}
Density functional theory calculations were performed using the Vienna \textit{ab initio} simulation package (VASP) \cite{kresse1993ab,kresse1996efficiency}. The Perdew-Burke-Ernzerhof (PBE) functional \cite{perdew1996generalized} was used for structural relaxation, with parameters consistent with those in the Materials Project \cite{jain2013commentary}. Specifically, a plane-wave energy cutoff of 520 eV was applied, and the Brillouin zone was integrated using a $k$-point density of 100 per reciprocal atom volume. Phase stability was evaluated based on the energy above the hull ($E_{\rm hull}$)\cite{ongspLi}, which is the decomposition energy of the target
compound relative to the most stable competing phases on the phase diagram, as schematically illustrated in Figure S1. Phonon spectrum calculations were performed using the finite displacement method via Phonopy \cite{togo2023implementation}. The electronic energy and force convergence criteria in phonon spectrum calculations were set to 10$^{-8}$ eV and 10$^{-3}$ eV/\r{A}, respectively.

Additionally, meta-GGA functionals, including SCAN \cite{sun2015strongly} and R2SCAN \cite{furness2020accurate}, as well as the random-phase approximation (RPA) \cite{harl2010assessing}, were employed for phase stability comparison. For meta-GGA calculations, the energy and force convergence criteria were set to 10$^{-5}$ eV and 10$^{-2}$ eV/\r{A}, respectively, with a $k$-point density matching that used in the PBE calculations. In RPA single-point calculations, the SCAN-optimized structure was used with an energy cutoff of 600 eV. The energy convergence criterion was 10$^{-8}$ eV, and a $k$-point line density of 0.25 \r{A}$^{-1}$ was applied. All structure manipulation and analysis were performed using Pymatgen\cite{ong2013python}.

\section{Results}
\subsection{Benchmark: Rediscovering Known Materials}
To evaluate the effectiveness of uMLIP-driven crystal structure prediction, we benchmarked its performance by attempting to rediscover two known quaternary compounds: \ce{Sr2LiAlO4} ($P2_1/m$) \cite{wang2018mining} and \ce{Ba2YAlO5} ($P2_1/m$)\cite{simura2023perovskite}. These systems were specifically chosen because they were not part of the uMLIP's training set, providing a direct test of the model's extrapolation capability for new chemical systems.

\begin{figure}[h]
    \centering
    \includegraphics[width=0.85\textwidth]{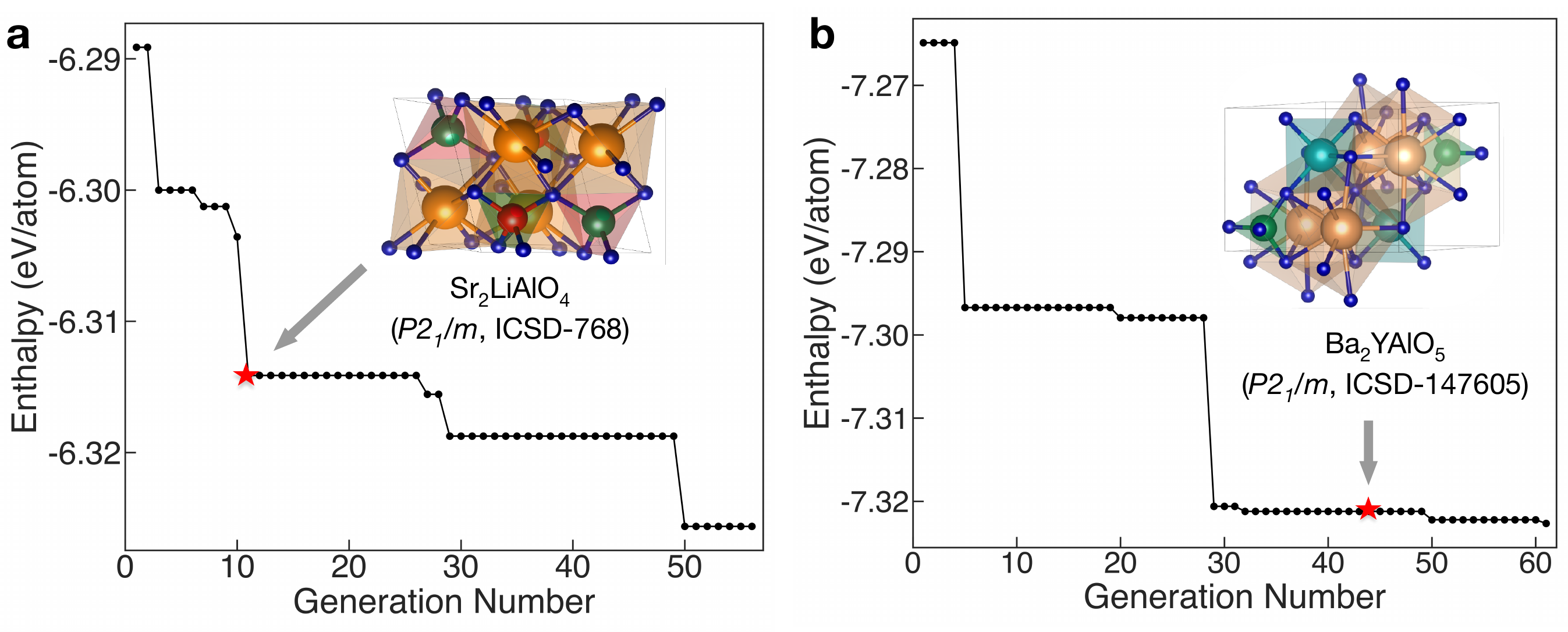}
    \caption{Evolution of the lowest formation enthalpy as a function of generation during the crystal structure prediction searches for (a) \ce{Sr2LiAlO4} and (b) \ce{Ba2YAlO5}. The red star marks the generation in which the correct experimental structure, shown in the inset, was first identified.}
    \label{fig:benchmark}
\end{figure}

As shown in Figure~\ref{fig:benchmark}, our benchmark searches successfully identified the experimentally known structures for both target compounds. The known monoclinic structure of \ce{Sr2LiAlO4} was found in the 11th generation, while the correct monoclinic structure for \ce{Ba2YAlO5} emerged in the 44th generation. The consistent decrease in the calculated enthalpy throughout both searches, as depicted in the figure, demonstrates the effectiveness of the uMLIP in guiding the evolutionary algorithm toward the global minimum energy structure.

\subsection{New Materials in Sr-Li-Al-O Chemistry}
\begin{figure}[H]
    \centering
    \includegraphics[width=0.75\textwidth]{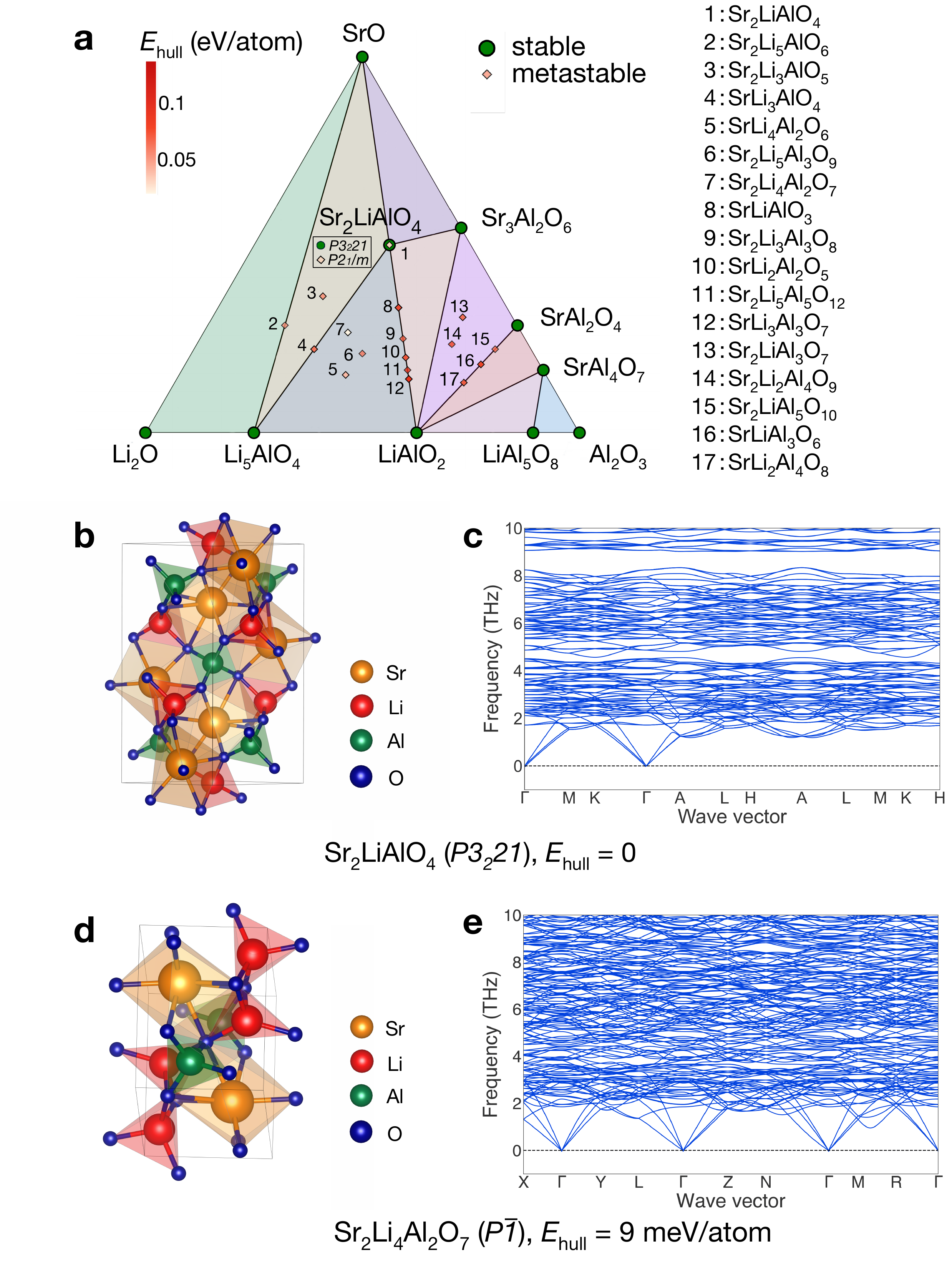}
    \caption{(a) Calculated phase diagram of the Sr-Li-Al-O system. The green solid circle represents the stable phase, while colored diamonds denote metastable phases. The lighter the color of the diamond, the more unstable the phase. (b-e) Crystal structures and corresponding calculated phonon spectra for two newly identified stable compounds, \ce{Sr2LiAlO4} and \ce{Sr2Li4Al2O7}.}
    \label{fig:slao_pd}
\end{figure}

Using uMLIP-driven crystal structure prediction, we investigated 17 distinct compositions within the Sr–Li–Al–O chemical system. Figure \ref{fig:slao_pd}a presents the calculated phase diagram for this quaternary system, while Table \ref{tab:table1} and Table S1 summarize the calculated properties of the newly identified materials. Among them, we identified six new compounds with $E_{\rm hull}$ below 30 meV/atom. This threshold was selected based on previous benchmark studies\cite{wang2018mining} indicating that materials with $E_{\rm hull}$ below this value have a high probability of experimental synthesizability. 

Notably, two of these materials have $E_{\rm hull}$ lower than that of the experimentally known \ce{Sr2LiAlO4} in its monoclinic $P2_1/m$ phase. One is a novel polymorph of \ce{Sr2LiAlO4}, crystallizing in a trigonal structure with space group $P3_221$ and a $E_{\rm hull}$ of 0 meV/atom. (Figure \ref{fig:slao_pd}b) The other is \ce{Sr2Li4Al2O7}, which adopts a triclinic structure ($P\bar{1}$) with a $E_{\rm hull}$ of 9 meV/atom. (Figure \ref{fig:slao_pd}d) 

The trigonal polymorph of \ce{Sr2LiAlO4} has lattice parameters of $a$ = 5.61 \AA, $c$ = 12.80 \AA, with angles $\alpha$ = 90°, $\beta$ = 90°, and $\gamma$ = 120°. Two experimental polymorphs of \ce{Sr2LiAlO4} have been reported: a monoclinic phase ($P2_1/m$) \cite{wang2018mining}, and a disordered orthorhombic phase ($Cmcm$) \cite{hoerder2019polymorphs}. The newly identified trigonal phase is 8 meV/atom more stable than the monoclinic form, indicating favorable thermodynamic stability (Table \ref{tab:table1}). Furthermore, phonon calculations reveal no imaginary frequencies across the Brillouin zone (Figure \ref{fig:slao_pd}c), confirming the dynamic stability and suggesting the phase has a high likelihood of being synthesizable.

The triclinic \ce{Sr2Li4Al2O7} structure has lattice parameters of $a$ = 5.25 \AA, $b$ = 5.45 \AA, $c$ = 6.72 \AA, with angles $\alpha$ = 102.19°, $\beta$ = 94.87°, and $\gamma$ = 108.15°. This compound is 1 meV/atom less stable than the monoclinic \ce{Sr2LiAlO4} phase, further indicating good phase stability. Interestingly, its calculated $E_{\rm hull}$ remains unchanged upon swapping Li and Al atomic positions, suggesting that this material may be disordered if synthesized. Its phonon spectrum also shows no imaginary modes, reinforcing its phase stability.(Figure \ref{fig:slao_pd}e)

\begin{table}[H]
\centering
\renewcommand{\arraystretch}{1.3}
\begin{tabular}{lcccc}
\hline
\textbf{Materials} & \textbf{Space Group} & \textbf{$E_{\rm hull}$} & \textbf{$N_{\rm atom}$} \\ \hline
\multicolumn{4}{l}{\textbf{Sr-Li-Al-O}} \\ \hline
\multirow{2}{*}{\ce{Sr2LiAlO4}}        & $P3_221$ (154)    & 0  & 24 \\ \
                     & $P2_1/m$ (11) $\star$    & 8  & 16 \\
\ce{Sr2Li4Al2O7}    & $P\overline{1}$ (2)   & 9  & 15 \\  
\multirow{2}{*}{\ce{SrLiAlO3}}        & $P4_3$ (78)   & 16  & 24 \\ 
                     & $C2/c$ (15)          & 25  & 12 \\ 
\ce{SrLi4Al2O6}     & $C2$ (5)          & 26 & 26 \\ 
\ce{Sr2Li3AlO5}     & $P1$ (1)          & 29 & 11 \\ \hline
\multicolumn{4}{l}{\textbf{Ba-Y-Al-O}} \\ \hline
\ce{Ba2YAlO5} (exp.)       & $P2_1/m$ (11) $\star$        & 0  & 18 \\   
\ce{Ba4YAlO7}       & $P1$ (1)          & 28 & 26 \\ \hline
\end{tabular}
\caption{Predicted novel materials in the Sr-Li-Al-O and Ba-Y-Al-O chemical systems, sorted by phase stability ($E_{\rm hull}$ in meV/atom). The table lists the chemical formula, space group, and the number of atoms ($N_{\rm atom}$) in the primitive cell for each entry. Known experimental phases are marked with a star symbol for comparison.}
\label{tab:table1}
\end{table}

In addition to these two compounds, we identified four other metastable yet promising phases. For \ce{SrLiAlO3}, two polymorphs were found: a tetragonal phase ($P4_3$) with $E_{\rm hull}$ = 16 meV/atom and a monoclinic phase ($C2/c$) with $E_{\rm hull}$ = 25 meV/atom. The monoclinic phase likely exhibits Li/Al disorder, as its energy remains unchanged upon site swapping. The remaining compounds—monoclinic \ce{SrLi4Al2O6} ($E_{\rm hull}$ = 26 meV/atom) and triclinic \ce{Sr2Li3AlO5} ($E_{\rm hull}$ = 29 meV/atom)—are also metastable. Despite their relatively higher calculated $E_{\rm hull}$ values, phonon calculations show that all four phases are dynamically stable, indicating potential for experimental realization. (Figure S2) 

\subsection{New Materials in Ba-Y-Al-O Chemistry}
\begin{figure}
    \centering
    \includegraphics[width=0.7\textwidth]{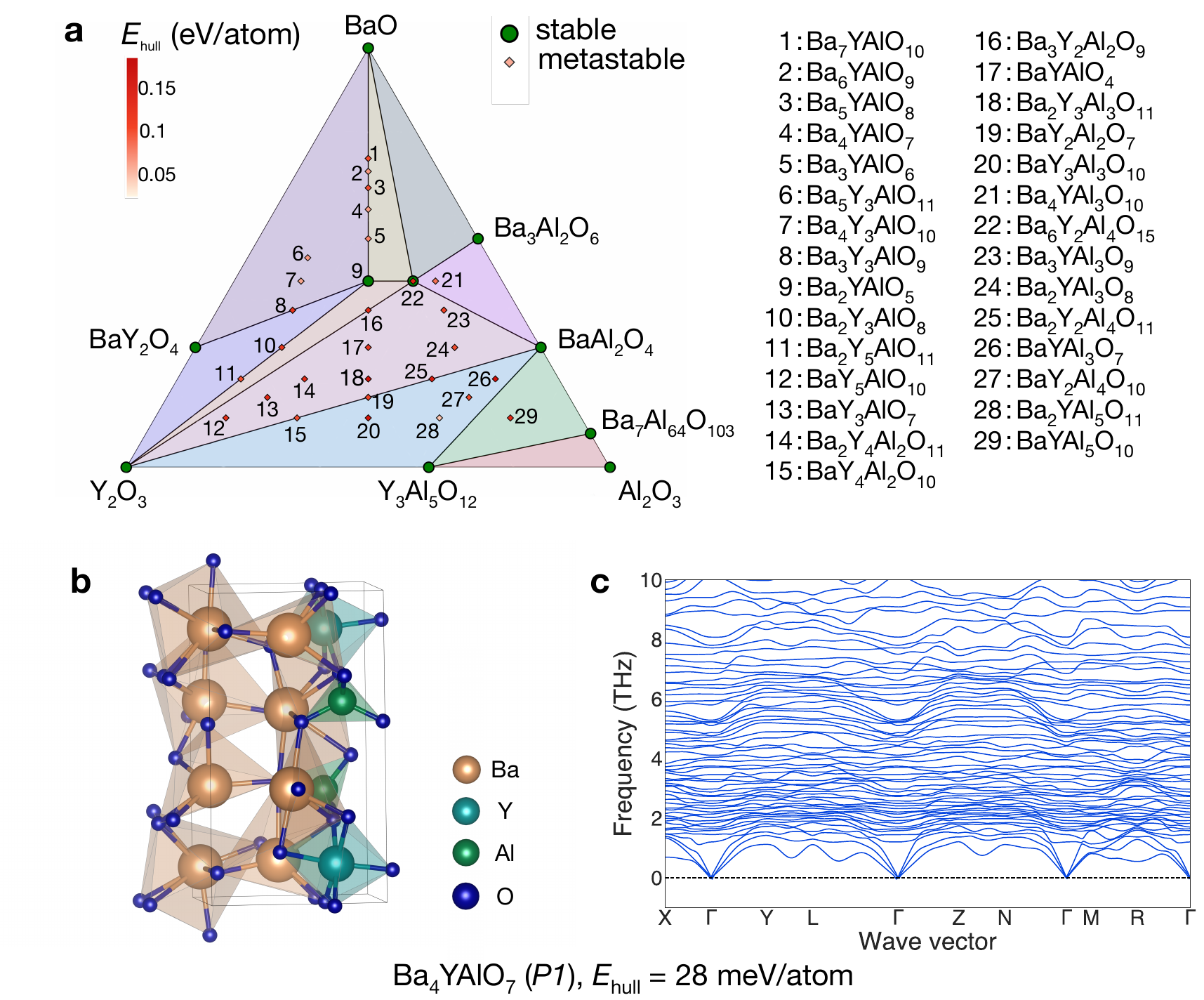}
    \caption{(a) Calculated phase diagram of the Ba–Y–Al–O system. The green solid circle represents a stable phase, while colored diamonds denote metastable phases. The lighter the color of the diamond, the less stable the phase. (b-c) Crystal structure and corresponding calculated phonon spectrum of \ce{Ba4YAlO7}.}
    \label{fig:byao_pd}
\end{figure}

In a similar vein, we thoroughly explored 29 distinct compositions within the Ba–Y–Al–O chemical system. Figure \ref{fig:byao_pd}a displays the calculated phase diagram for this quaternary system, while Table \ref{tab:table1} and Table S2 summarize the calculated properties of the newly identified materials. In contrast to the Sr–Li–Al–O system, only two compounds were found with $E_{\rm hull}$ below 30 meV/atom. One of these is the experimentally known compound \ce{Ba2YAlO5} \cite{simura2023perovskite}, which has an $E_{\rm hull}$ of 0 meV/atom, indicating it is a stable phase in the calculated phase diagram. The other compound, \ce{Ba4YAlO7}, crystallizes in a triclinic structure (space group $P1$) and is metastable, with an $E_{\rm hull}$ of 28 meV/atom. Phonon spectrum calculations for \ce{Ba4YAlO7} reveal no imaginary frequencies, confirming its dynamical stability.

\section{Discussion}
In this work, we identified a new, potentially more stable phase ($P3_221$) of \ce{Sr2LiAlO4}, in comparison to the experimentally reported $P2_1/m$ phase. Although this discovery is intriguing, it contradicts experimental observations. The monoclinic \ce{Sr2LiAlO4} ($P2_1/m$) is the first successfully synthesized quaternary compound in the Sr-Li-Al-O chemical system\cite{wang2018mining}, suggesting it is likely the true ground state, as it would be the most thermodynamically favorable phase to form.

\begin{figure}[h]
    \centering    
    \includegraphics[width=0.6\textwidth]{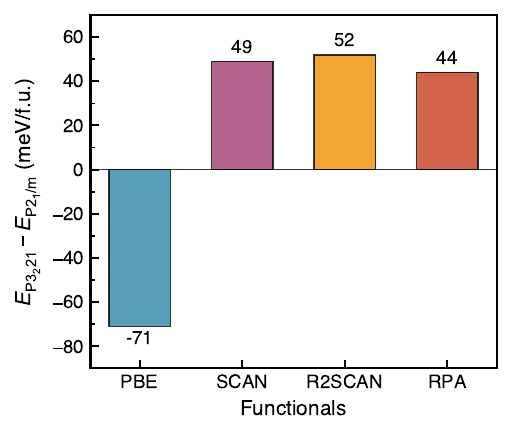}
    \caption{Calculated energy differences between the \ce{Sr2LiAlO4} polymorphs ($P2_1/m$ and $P3_221$) using various exchange-correlation functionals.}
    \label{fig:xc}
\end{figure}

To resolve this discrepancy, we evaluated the relative phase stability of the two \ce{Sr2LiAlO4} polymorphs using more accurate exchange-correlation functionals. As shown in Figure~\ref{fig:xc}, we employed higher-level methods, including SCAN, R2SCAN, and the Random Phase Approximation (RPA). Both SCAN and R2SCAN predict the experimental $P2_1/m$ phase to be the ground state, with energies 49 meV/f.u. and 52 meV/f.u. lower, respectively, than that of the $P3_221$ phase. The RPA calculation—widely regarded as the gold standard for phase stability evaluation—corroborates these results, showing the experimental phase to be more stable by 44 meV/f.u. In stark contrast, the PBE functional incorrectly predicts the $P3_221$ phase to be more stable by 71 meV/f.u.

These results strongly indicate that the experimental $P2_1/m$ phase is indeed the true ground state, while the $P3_221$ structure is metastable. More broadly, this finding has important implications for the current generation of uMLIPs. Since most uMLIPs are trained on large datasets computed with the PBE functional, they may inherit its systematic errors—such as incorrectly predicting the phase stability of materials. Our results thus underscore the need for caution when using such uMLIPs to predict novel stable materials for experimental validation.

As previously discussed, a primary advantage of uMLIPs in CSP is their role as highly efficient surrogate models for first-principles calculations like DFT. It is widely acknowledged that uMLIPs can predict the energy and forces of a structure at speeds at least three orders of magnitude faster than DFT.\cite{zuo2020performance} To quantify this efficiency gain, we conducted a direct comparison for the 15-atom \ce{Sr2Li4Al2O7} system. Identifying its most stable structure using DFT-CSP required 1467 CPU-hours, whereas the M3GNet-driven CSP took only 37 CPU-hours, representing a speedup factor of approximately 40 (Table S4). This performance advantage becomes even more pronounced for larger systems, as the computational cost of DFT (based on the PBE functional) scales cubically with the number of atoms ($O(N^3)$)). This speedup enables large-scale explorations, such as those performed in this work, that would be computationally prohibitive using traditional methods. 

We also examined the experimentally known phase \ce{Ba6Y2Al4O15} within the Ba-Y-Al-O chemical system. This compound contains 54 atoms in its primitive cell and is included in the M3GNet training data. Despite our efforts, we were unable to reproduce this phase using our approach. The primary reason for this failure likely lies not in the energetic inaccuracy of the uMLIP, but in the inefficiency of the evolutionary algorithm used to navigate such a vast structural search space. Historically, the inefficiency of CSP has been attributed to the high cost of DFT calculations. However, our findings suggest that with the advent of modern surrogate models, the bottleneck has shifted to the CSP algorithm itself. Therefore, future research should focus on developing more efficient CSP algorithms for structure prediction.

\section{Conclusion}
In summary, we have demonstrated the capability of uMLIPs, specifically M3GNet, to accelerate CSP for complex quaternary oxides that are otherwise inaccessible to traditional DFT-based CSP. By exploring the Sr–Li–Al–O and Ba–Y–Al–O chemical systems, we successfully rediscovered known materials excluded from the training data and identified seven new thermodynamically and dynamically stable compounds, including a novel trigonal phase of \ce{Sr2LiAlO4} and a disordered \ce{Sr2Li4Al2O7}. However, our results also reveal limitations in current uMLIP models, particularly their reliance on PBE-based training data, which can lead to incorrect phase stability predictions. Benchmarking with higher-level methods such as SCAN and RPA underscores the necessity of cross-validation for reliable experimental guidance. Moreover, the main bottleneck in CSP has shifted from computational cost to the efficiency of structural search algorithms. These findings highlight both the promise and the challenges of uMLIP-driven CSP in the discovery of complex materials.

\section{CRediT authorship contribution statement}
\textbf{Yuqi An:} Writing – original draft, Investigation, Formal analysis, Data curation, Methodology. \textbf{Zhenbin Wang:} Writing – review \& editing, Supervision, Resources, Funding acquisition, Conceptualization, Formal analysis.

\section{Declaration of competing interest}
The authors declare no competing interests.

\begin{acknowledgement}
This work was financially supported by the City University of Hong Kong Start-up Grant (9020004). Some of the calculations were performed using the computational facilities of CityU Burgundy, which are managed and provided by the Computing Services Centre at the City University of Hong Kong.

\end{acknowledgement}

\section{Appendix A. Supplementary data}

Supplementary data to this article can be found on ESI. 

\section{Data availability}
All discovered structures are available on \href{https://figshare.com/articles/dataset/New_structures_of_Sr-Li-Al-O_and_Ba-Y-Al-O_quaternary_compounds/29652701}{Figshare}.

% This is where you add your references. Uncomment the line below and point it to your bibtex file.
%\bibliography{refs}

\clearpage

\bibliography{ref.bib}
\end{document}

\begin{figure}[htbp]
    \centering
    \includegraphics[width=0.6\textwidth]{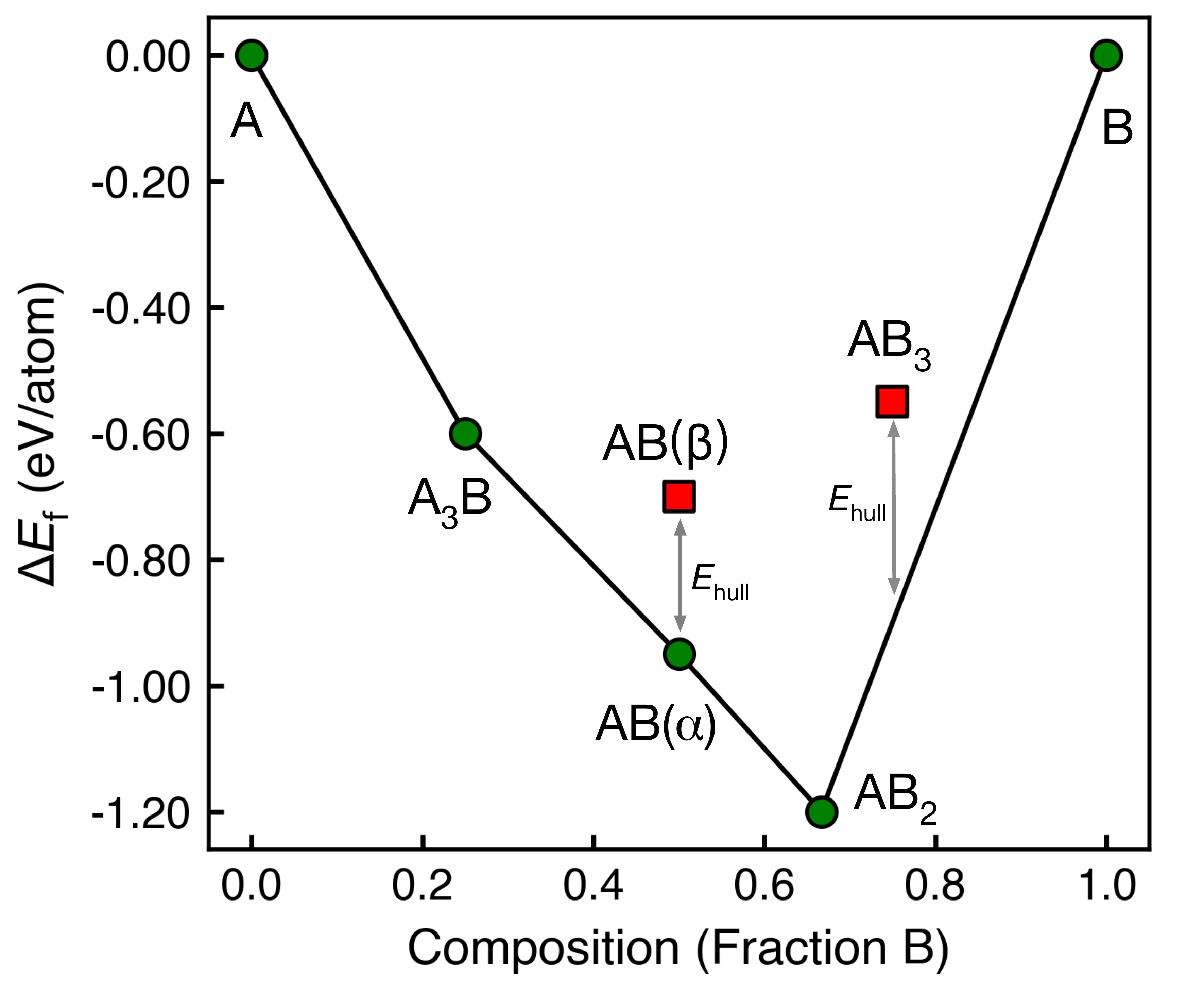}
    \caption{Calculated phase diagram for the A-B binary system. $\Delta$$E_{\rm f}$ is the computed formation energy. Green circles represent stable phases lying on the convex hull (solid black line), while red squares show metastable phases above the hull. The vertical double-arrows indicate the energy above hull ($E_{\rm hull}$) for metastable phases. Stable phases are \ce{A}, \ce{A3B}, \ce{AB}($\alpha$), \ce{AB2}, and \ce{B}, metastable phases are \ce{AB}($\beta$) and \ce{AB3}.}
    \label{fig:ehull}
\end{figure}

\begin{figure}[htbp]
    \centering
    \includegraphics[width=0.6\textwidth]{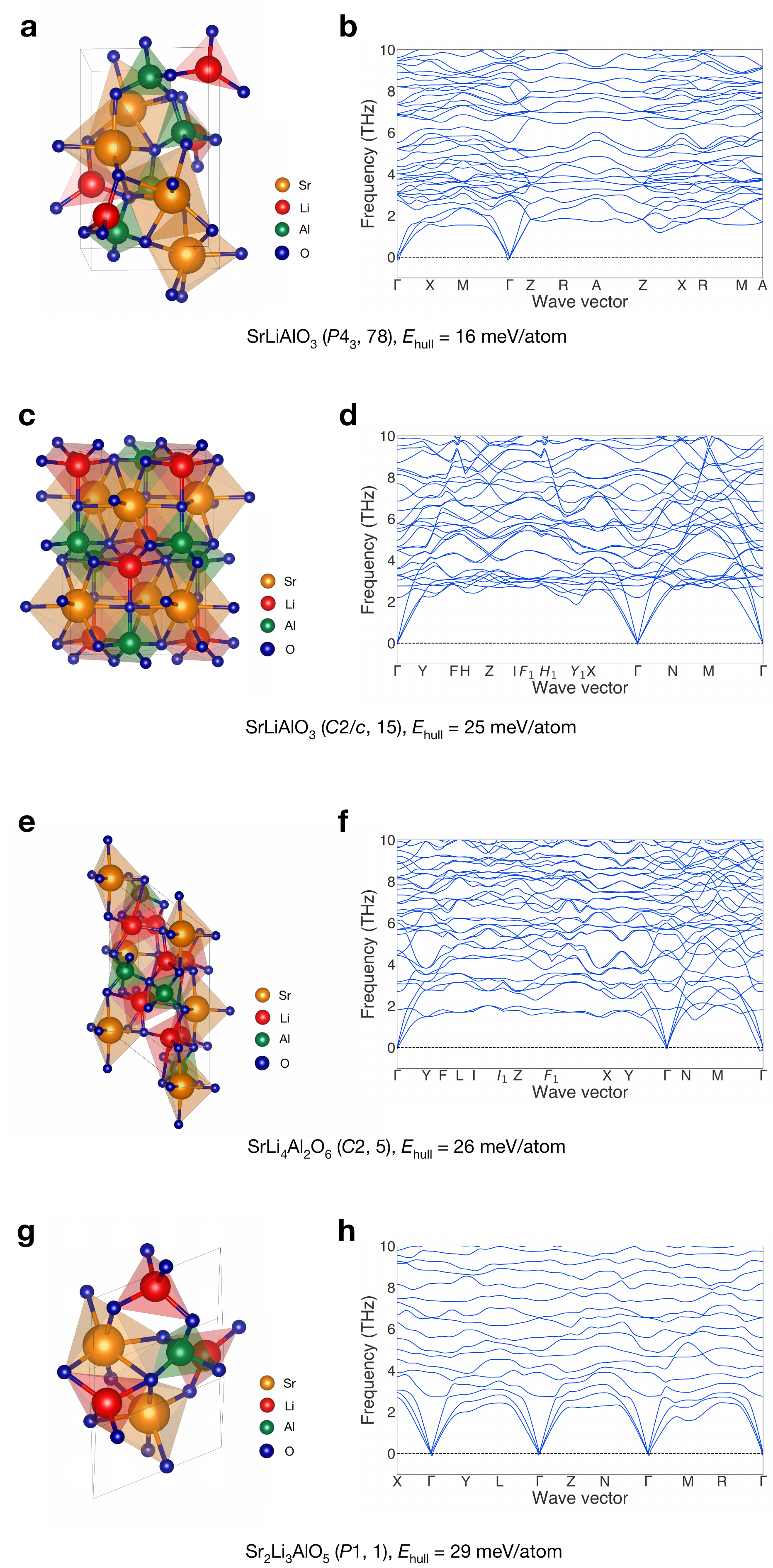}
    \caption{Calculated crystal structures, space groups, phase stability ($E_{\rm hull}$), and phonon spectra for four newly identified metastable compounds: (a-b) \ce{SrLiAlO3} ($P4_3$) (c-d) \ce{SrLiAlO3} ($C2/c$) (e-f) \ce{SrLi4Al2O6} ($C2$) and (g-h) \ce{Sr2Li3AlO5} ($P1$).}
    \label{fig:slao_si}
\end{figure}

\clearpage

\begin{table}[ht]
\centering
\renewcommand{\arraystretch}{1.2} 
\begin{tabular}{lccc}
\hline
\textbf{\makecell[l]{Materials}} & \textbf{\makecell[l]{Space Group}} & \textbf{\makecell{$E_{\rm hull}$}} & \textbf{\makecell{$N_{\rm atom}$}} \\ \hline
\multirow{3}{*}{\ce{SrLiAlO3}}  
                      & $P1$ (1)             & 37  & 18 \\
                      & $Imm2$ (44)          & 82  & 6  \\
                      & $P1$ (1)             & 93  & 30 \\ 
\multirow{3}{*}{\ce{SrLi3AlO4}}        & $P1$ (1)             & 48  & 18 \\
                      & $Cm$ (8)             & 45  & 9  \\
                      & $P1$ (1)             & 59  & 27 \\  
\ce{Sr2Li3AlO5}       & $P\overline{4}2_1m$ (113) & 46  & 22 \\ 
\ce{Sr2Li5AlO6}       & $P1$ (1)             & 47  & 14 \\ 
\ce{SrLi4Al2O6}       & $P1$ (1)             & 51  & 13 \\ 
\multirow{2}{*}{\ce{Sr2LiAl3O7}}       & $P1$ (1)             & 55  & 26 \\
                      & $Cm$ (8)             & 93  & 13 \\ 
\ce{Sr2Li5Al3O9}      & $P1$ (1)             & 56  & 19 \\ 
\ce{Sr2LiAl5O10}      & $P1$ (1)             & 62  & 18 \\ 
\multirow{2}{*}{\ce{SrLi2Al4O8}}       & $P1$ (1)             & 63  & 15 \\
                      & $P1$ (1)             & 91  & 30 \\ 
\multirow{2}{*}{\ce{SrLi2Al2O5}}       & $P2_1$ (4)           & 67  & 20 \\
                      & $P1$ (1)             & 74  & 10 \\ 
\ce{Sr2Li3Al3O8}      & $P1$ (1)             & 68  & 16 \\ 
\multirow{2}{*}{\ce{SrLiAl3O6}}        & $P1$ (1)             & 74  & 22 \\
                      & $P1$ (1)             & 80  & 11 \\ 
\ce{Sr2Li2Al4O9}      & $P1$ (1)             & 78  & 17 \\ 
\ce{Sr2LiAlO4}        & $Pmm2$ (25)          & 82  & 8  \\
\multirow{2}{*}{\ce{SrLi3Al3O7}}       & $Imm2$ (44)          & 88  & 28 \\
                      & $Cm$ (8)             & 88  & 14 \\ 
\ce{Sr2Li5Al5O12}     & $P1$ (1)             & 95  & 24 \\ 
\hline
\end{tabular}
\caption{Predicted novel materials in the Sr-Li-Al-O chemical systems, sorted by phase stability ($E_{\rm hull}$ in meV/atom). The table lists the chemical formula, space group, and the number of atoms ($N_{\rm atom}$) in the primitive cell for each entry.}
\label{tab:SLAO_moredetails}
\end{table}

\clearpage
\renewcommand{\arraystretch}{0.8}
\begin{longtable}{lccc}
\caption{Predicted novel materials in the Ba-Y-Al-O chemical systems, sorted by phase stability ($E_{\rm hull}$ in meV/atom). The table lists the chemical formula, space group, and the number of atoms ($N_{\rm atom}$) in the primitive cell for each entry.} 
\label{tab:BYAO_corrected} \\
\hline
\textbf{\makecell[l]{Materials}} & \textbf{\makecell[l]{Space Group}} & \textbf{\makecell{$E_{\rm hull}$}} & \textbf{\makecell{$N_{\rm atom}$}} \\ \hline
\endfirsthead

\multicolumn{4}{c}%
{{\bfseries \tablename\ \thetable{} -- continued from previous page}} \\
\hline
\textbf{\makecell[l]{Materials}} & \textbf{\makecell[l]{Space Group}} & \textbf{\makecell{$E_{\rm hull}$}} & \textbf{\makecell{$N_{\rm atom}$}} \\ \hline
\endhead

\hline \multicolumn{4}{r}{{Continued on next page}} \\ \hline
\endfoot

\hline
\endlastfoot

\ce{BaY2Al4O10}      & $P2$ (3)                & 36  & 17 \\ 
\ce{Ba4YAl3O10}      & $P1$ (1)                & 37  & 18 \\ 
\ce{Ba5Y3AlO11}      & $Cm$ (8)                & 48  & 20 \\ 
\multirow{2}{*}{\ce{Ba5YAlO8}}        & $P\overline{1}$ (2)             & 50  & 30 \\ 
                     & $P1$ (1)                & 57  & 15 \\ 
\ce{Ba6YAlO9}        & $P1$ (1)                & 55  & 17 \\ 
\multirow{2}{*}{\ce{BaYAl3O7}}        & $P1$ (1)                & 57  & 24 \\ 
                     & $P312$ (149)            & 72  & 12 \\ 
\multirow{3}{*}{\ce{BaYAlO4}}         & $P2_1/m$ (11)           & 58  & 14 \\
                     & $P1$ (1)                & 89  & 21 \\  
                     & $Cm$ (8)                & 156 & 7 \\  
\multirow{2}{*}{\ce{Ba2YAl3O8}}       & $P1$ (1)                & 59  & 14 \\
                     & $P1$ (1)                & 94  & 28 \\ 
\ce{Ba4Y3AlO10}      & $P1$ (1)                & 61  & 18 \\
\ce{Ba3YAl3O9}       & $P1$ (1)                & 61  & 16 \\ 
\ce{Ba4YAlO7}        & $P1$ (1)                & 69  & 13 \\  
\multirow{2}{*}{\ce{Ba3YAlO6}}        & $R\overline{3}c$ (167)  & 72  & 22 \\
                     & $P1$ (1)                & 131 & 11 \\  
\ce{Ba2Y2Al4O11}     & $P\overline{1}$ (2)                & 76  & 19 \\
\ce{BaY4Al2O10}      & $C2$ (5)                & 81  & 17 \\
\ce{Ba7YAlO10}       & $P1$ (1)                & 88  & 19 \\
\ce{Ba3YAl3O9}        & $P1$ (1)               & 94  & 32 \\ 
\ce{Ba2Y4Al2O11}     & $P1$ (1)                & 95  & 19 \\ 
\multirow{2}{*}{\ce{BaY3AlO7}}        & $Cm$ (8)                & 99  & 12 \\
                     & $P\overline{1}$ (2)     & 133 & 24 \\ 
\ce{BaY5AlO10}       & $Cm$ (8)                & 101 & 17 \\
\ce{Ba2Y5AlO11}      & $P1$ (1)                & 101 & 19 \\ 
\ce{Ba3Y3AlO9}       & $P1$ (1)                & 102 & 16 \\ 
\ce{BaY2Al2O7}       & $C2$ (5)                & 104 & 12 \\
\multirow{2}{*}{\ce{Ba2Y3AlO8}}       & $P1$ (1)                & 105 & 28 \\ 
                     & $Cm$ (8)                & 108 & 14 \\                  
\ce{BaYAl5O10}       & $P1$ (1)                & 108 & 17 \\ 
\ce{Ba2YAl5O11}      & $P1$ (1)                & 111 & 19 \\ 
\ce{Ba2YAlO5}        & $P1$ (1)                & 118 & 9 \\  
\ce{Ba6Y2Al4O15}     & $P1$ (1)                & 120 & 27 \\
\ce{Ba2Y3Al3O11}     & $P1$ (1)                & 135 & 19 \\ 
\ce{BaY3Al3O10}      & $P1$ (1)                & 138 & 17 \\ 
\ce{Ba3Y2Al2O9}      & $P\overline{6}m2$ (187) & 139 & 16 \\  
\end{longtable}

\begin{table}[h]
\centering
\caption{Comparison of predicted materials for 12-atom \ce{SrLiAlO3} using DIRECT-sampled and original M3GNet models. The table shows space group, phase stability ($E_{\text{hull}}$ in meV/atom), lattice parameters, and volume per unit cell for the most stable phase predicted by each model. Lattice angles $\alpha$ and $\gamma$ are 90° for both structures; only the non-90° angle $\beta$ is reported.}
\renewcommand{\arraystretch}{1.5}
\label{tab:m3gnet_comparison}
\begin{tabular}{lcccc}
\hline
\textbf{Model} & \textbf{Space Group} & \textbf{$E_{\rm hull}$} & \textbf{Lattice Parameter} & \textbf{Volume} \\
\hline
M3GNet-DIRECT & $C2/c$ (15) & 25 & \begin{tabular}[c]{@{}c@{}}a = 5.49, b = 9.86 \\ c = 5.16, $\beta$ = $94.53^\circ$\end{tabular} & 278.38 \\  
M3GNet & $P2_1$ (4) & 31 & \begin{tabular}[c]{@{}c@{}}a = 5.79, b = 5.19 \\ c = 5.96, $\beta$ = $117.05^\circ$\end{tabular} & 159.47 \\
\hline
\end{tabular}
\end{table}

\begin{table}[h]
\centering
\caption{Computational efficiency comparison between DFT and uMLIP methods for \ce{Sr2Li4Al2O7} CSP. The table lists the total computational time for complete CSP search and speedup factor relative to the DFT baseline.}
\renewcommand{\arraystretch}{1.5}
\label{tab:efficiency_comparison}
\begin{tabular}{lcc}
\hline
\textbf{Approach} & \textbf{CPU Time (hour)} & \textbf{Speedup Factor} \\
\hline
DFT & 1,467 & 1.0× (baseline) \\
uMLIP & 37 & \textbf{40×} \\
\hline
\end{tabular}
\end{table}